\documentclass[%
 reprint,
 superscriptaddress,
 amsmath,
 amssymb,
 aps, 
 prl
]{revtex4-2}

\usepackage{graphicx}
\usepackage{dcolumn}
\usepackage{bm}
\usepackage{booktabs}
\usepackage{makecell}
\usepackage{multirow}
\usepackage[caption=false]{subfig}
\usepackage[colorlinks=true, linkcolor=blue, urlcolor=blue, citecolor=blue]{hyperref}
\usepackage{upgreek}

\begin{document}

\preprint{}

\title{\textbf{Spin-orbit magnetism in altermagnets} 
}%

\author{Ruojia Wang}
\affiliation{State Key Laboratory of Quantum Functional Materials, Department of Physics, and Guangdong Basic Research Center of Excellence for Quantum Science, Southern University of Science and Technology (SUSTech), Shenzhen 518055, China}
\author{Yuntian Liu}
\affiliation{State Key Laboratory of Quantum Functional Materials, Department of Physics, and Guangdong Basic Research Center of Excellence for Quantum Science, Southern University of Science and Technology (SUSTech), Shenzhen 518055, China}
\author{Renzheng Xiong}
\affiliation{State Key Laboratory of Quantum Functional Materials, Department of Physics, and Guangdong Basic Research Center of Excellence for Quantum Science, Southern University of Science and Technology (SUSTech), Shenzhen 518055, China}
\author{Xiaobing Chen}
\email{chenxiaobing@quantumsc.cn}
\affiliation{Quantum Science Center of Guangdong–Hong Kong–Macao Greater Bay Area (Guangdong), Shenzhen 518045, China}
\author{Qihang Liu}
\email{liuqh@sustech.edu.cn}
\affiliation{State Key Laboratory of Quantum Functional Materials, Department of Physics, and Guangdong Basic Research Center of Excellence for Quantum Science, Southern University of Science and Technology (SUSTech), Shenzhen 518055, China}
\affiliation{Quantum Science Center of Guangdong–Hong Kong–Macao Greater Bay Area (Guangdong), Shenzhen 518045, China}

\date{\today}

\begin{abstract}
While the anomalous Hall effect was proposed as a transport fingerprint of altermagnets, it originates from the spin-orbit coupling-induced net magnetization, corresponding to a distinct magnetic phase termed spin-orbit magnetism. However, the microscopic mechanism enabling spin-orbit magnets to yield a prominent anomalous Hall response despite a vanishingly small net magnetization remains elusive. Here, by employing oriented spin group theory and spin-orbit-coupling tensor expansion, we systematically disentangle the perturbative behaviors of orbital and spin magnetizations with respect to spin-orbit coupling in altermagnets. Remarkably, we find that only if the opposite-spin sublattices are connected through a fourfold rotation, the orbital and spin magnetizations exhibit distinct perturbative orders. In these altermagnets, we further discover a coaxial Hall effect characterized by the induced spin and orbital magnetizations aligning parallel to the N\'{e}el vector, which we further demonstrate by first-principles calculations in the altermagnet $\text{KV}_2\text{Se}_2\text{O}$. This effect holds great promise for achieving deterministic switching of the N\'{e}el order under weak external fields. Our work provides a systematic symmetry approach to identify potential altermagnetic candidates combining a large anomalous Hall effect with minimal net magnetization, paving the way for high-performance, stray-field-free spintronic applications.
\end{abstract}

\maketitle

Unconventional magnets, which combine the compensated magnetization with ferromagnetic-like physical properties, emerge as a pivotal platform for next-generation spintronics \cite{liu2025, chen2026rise}. They comprise two representative branches: anomalous Hall antiferromagnets \cite{Chen2014, Nakatsuji2015, RN17} and spin-splitting antiferromagnets \cite{Sandratskii1981, hayami-2019, Yuan2020, ma-2021, Igor2021}. One prototypical class of the latter is the altermagnet \cite{smejkal-2022PRX}, which is a collinear antiferromagnet that exhibits momentum-dependent spin splitting even in the absence of spin-orbit coupling (SOC). Although the anomalous Hall effect (AHE) was initially perceived as the transport fingerprint of altermagnets, recent insights have revealed that this effect in antiferromagnets corresponds to a distinct magnetic phase termed spin-orbit magnetism (SOM), in which a net magnetization arises exclusively from SOC as a consequence of its symmetry-breaking pathway \cite{Liu2026}. This SOC-induced magnetization, comprising both spin and orbital contributions, underpins the AHE, indicating that the AHE in altermagnets is fundamentally a manifestation of SOM \cite{Bai2025, solovyev-2026}. Notably, the net magnetization in altermagnet is generally weak yet varies by several orders of magnitude from $10^{-6}$ to $10^{-2}$ $\mu_B$ across different systems \cite{smejkal-2020, Igor2021, Gonzalez2023, Fakhredine2023, Lei2024, Attias2024, takagi-2024}, leaving the underlying connection between the SOC-induced AHE and the net magnetization largely unestablished \cite{smejkal-2020,liu-2023}. 

Practically, an altermagnet whose Néel vector orientation simultaneously places it in the SOM regime represents an optimal configuration: it generates a measurable AHE signal encoding the Néel order while exhibiting spin-polarized band structure \cite{Gonzalez2023, Lei2024, reichlova2024}. Such a synergy between altermagnetism and SOM, however, requires a precise understanding of how the SOC-induced magnetization depends on the SOC strength and on the Néel vector orientation. To elucidate this, symmetry-based approaches have been adopted to classify diverse anomalous Hall Néel textures \cite{Xiao2025}. However, conventional magnetic group analysis naturally entangles the contributions from magnetic order and SOC \cite{Brinkman1966, Litvin1977, sandratskii-1979, Liu2022}, obscuring the microscopic mechanism of the SOC-induced AHE in altermagnets. This limitation is fundamentally overcome by the recent development of spin space group theory \cite{Chen2024,Xiao2024,Jiang2024,Watanabe2024}. Within this framework, recent studies have employed representation analysis to study the relationships between the Néel order and the net magnetization \cite{McClarty2024, Roig2025, Schiff2025}. Furthermore, the spin-orbit tensor expansion method has enabled the resolution of the net magnetization into its constituent spin and orbital parts \cite{Liu-2025, Liu2026}, where the latter has the same symmetry constraints as AHE, successfully explaining their distinct perturbative orders with respect to SOC in ferromagnets \cite{Liu-2025} and the coplanar antiferromagnet Mn$_{3}$Sn \cite{Liu2026}. Nevertheless, a comprehensive framework that systematically quantifies SOC effects on both spin and orbital magnetizations, alongside their dependence on the Néel vector orientation, remains absent for altermagnets. This hinders the symmetry-guided exploration of potential altermagnets that simultaneously exhibit both a suppressed net magnetization and a substantial AHE.

In this Letter, we employ the newly developed oriented spin space group theory, which unifies the spin space group and magnetic space group frameworks \cite{Liu2026}, to systematically investigate the SOM in altermagnets, i.e., the dependence of spin magnetization and orbital magnetization on the SOC effect in altermagnets. We find that when opposite-spin sublattices are connected by a sixfold rotation or multiple high-fold rotations, both spin and orbital magnetizations appear as third-order perturbations with respect to SOC. Conversely, they both follow first-order perturbations when connected solely by twofold rotations. Remarkably, only if the opposite-spin sublattices are connected via a fourfold rotation, the spin and orbital magnetization can exhibit distinct perturbative orders with respect to SOC. Intriguingly, such systems host a coaxial Hall effect where the spin magnetization, orbital magnetization, and Néel vector are mutually parallel to each other. We further screen 7 candidate materials and predict the existence of the coaxial Hall effect in KV$_{2}$Se$_{2}$O via first-principles calculations. This work establishes a quantitative relationship among the net magnetization, the AHE, and the Néel vector in altermagnets, providing explicit guidelines for the design of high-performance antiferromagnetic spintronic devices. 

\paragraph{SOC tensor expansion --}

To describe the transformation of SOC under spin group operations, we reformulate the SOC Hamiltonian using a 3×3 SOC tensor $\chi$:
\begin{equation}
    \hat{H}_{SOC} = \lambda \hat{\mathbf{L}}^T\chi\hat{\bm{\sigma}} = \lambda\sum_{ij}\chi_{ij}\hat{L}_i\hat{\sigma}_j,
\end{equation}
where $\lambda$, $\hat{L}$, $\hat{\sigma}$ denote the SOC strength, effective orbital angular momentum operator and spin operator, respectively. Under this framework, the spin magnetization $\mathbf{M}_S$ and orbital magnetization $\mathbf{M}_O$ transform under a spin group element \(\{U||R\}\) as $D(U)\mathbf{M}_S$ and det$(U)$det$(R)D(R)\mathbf{M}_O$, where $D(U)$ and $D(R)$ denote transformation matrices corresponding to spin-space operation $U$ and lattice-space operation $R$, respectively. 

Since spatial inversion and fractional translations leave both $\mathbf{M}_S$ and $\mathbf{M}_O$ invariant, their symmetry constraints are completely determined by 10 spin Laue groups \cite{smejkal-2022PRX}, which we further extend to the oriented spin Laue groups (OSLGs) to establish the relationships of these magnetizations with SOC and the N\'{e}el vector $\mathbf{N}$. Within this framework, while general spin groups possess a $SO(3)$ freedom between spin-space and lattice-space bases, collinear magnets require only a $SO(2)$ freedom to describe the relation between $\mathbf{N}$ and the lattice. Consequently, the SOC tensor $\chi$ can be expressed in terms of the orientation of $\mathbf{N}$, specified by the polar angle $\theta$ and azimuthal angle $\phi$:
\begin{equation}
    \chi(\theta,\phi) = 
    \begin{pmatrix}
        \cos\theta\cos\phi & -\sin\phi & \sin\theta\cos\phi \\
        \cos\theta\sin\phi & \cos\phi & \sin\theta\sin\phi \\
        -\sin\theta & 0 & \cos\theta
    \end{pmatrix}.
\end{equation}

By treating SOC as a perturbation, the emergent $\mathbf{M}_O$ and $\mathbf{M}_S$ can be expanded as a power series of $\chi(\theta,\phi)$:
\begin{equation}
\begin{split}
    \mathbf{M}_a = & \omega_{a}^{(0)}+\lambda\sum_{ij}\omega_{a,ij}^{(1)}\chi_{ij}(\theta,\phi) \\
    & + \lambda^{2}\sum_{ijkl}\omega_{a,ij,kl}^{(2)}\chi_{ij}(\theta,\phi)\chi_{kl}(\theta,\phi)+\dots,
\end{split}
\end{equation}
where $\omega^{(n)}$ is the (2n+1)th-order symmetry-adapted tensor. Detailed information about SOC tensor method is provided in the Supplementary Material S1 \cite{suppl}.

\paragraph{Distinct SOC scaling behaviors --}
By evaluating the leading order of $\omega^{(n)}$ for the orientation $(\theta,\phi)$ where OSLGs reduce to net-magnetization-allowed magnetic Laue groups (MLGs) under SOC, we classify ten altermagnetic OSLGs into three categories based on the leading orders of $\mathbf{M}_O$ and $\mathbf{M}_S$ in the SOC expansion (TABLE \ref{tab:w}). Crucially, these lowest non-vanishing orders remain invariant across all net-magnetization-allowed directions.
\begin{table}[htbp]
\centering
\caption{Classification of ten altermagnetic OSLGs based on the lowest non-vanishing orders of $\mathbf{M}_O$ and $\mathbf{M}_S$. While derived using OSLGs, spin Laue group notations are adopted as these leading orders remain invariant across all directions that allow the nonzero net magnetization.}
\label{tab:w}
\begin{ruledtabular}
\begin{tabular}{cccc}
\thead{Type} & \thead{Spin Laue group} & \thead{Order of $\mathbf{M}_O$} & \thead{Order of $\mathbf{M}_S$} \\
\midrule

\multirow{5}{*}{I} &
$^{-1}2 / \, ^{-1}m \, ^{\infty m}1$ &
\multirow{5}{*}{1} &
\multirow{5}{*}{1} \\

& $^{-1}m \, ^{-1}m \, ^{1}m \, ^{\infty m}1$ & & \\

& $^{1}4 / \, ^{1}m \, ^{-1}m \, ^{-1}m ^{\infty m}1$ & & \\


& $^{1} - 3 \, ^{-1}m ^{\infty m}1$ & & \\

& $^{1}6 / \, ^{1}m \, ^{-1}m \, ^{-1}m ^{\infty m}1$ & & \\
\midrule

\multirow{3}{*}{II} &
$^{-1}6 / \, ^{-1}m ^{\infty m}1$ &
\multirow{3}{*}{3} &
\multirow{3}{*}{3} \\

& $^{-1}6 / \, ^{-1}m \, ^{-1}m \, ^{1}m ^{\infty m}1$ & & \\

& $^{1}m \, ^{1} - 3 \, ^{-1}m ^{\infty m}1$ & & \\
\midrule

\multirow{2}{*}{III} &
$^{-1}4 / \, ^{1}m ^{\infty m}1$ &
\multirow{2}{*}{1} &
\multirow{2}{*}{2} \\

& $^{-1}4 / \, ^{1}m \, ^{-1}m \, ^{1}m ^{\infty m}1$ & & \\
\end{tabular}
\end{ruledtabular}
\end{table}

This classification is dictated by the spatial rotations connecting opposite-spin sublattices. For type I groups, where opposite-spin sublattices are connected solely by twofold rotations, both $\mathbf{M}_O$ and $\mathbf{M}_S$ emerge as first-order perturbations. This accounts for the relatively large net moments of $10^{-1}\,\mu_B$ theoretically predicted in $\text{RuF}_4$ \cite{Mil2024}. In type II groups, where opposite-spin sublattices are connected by sixfold or multiple high-fold rotations, both magnetizations behave as third-order perturbations. This is consistent with the small net moments of $10^{-5} \sim 10^{-4}\,\mu_B$ experimentally observed in type II altermagnets like $\text{FeS}$, $\text{Mn}_5\text{Si}_3$, and $\text{MnTe}$ \cite{Gonzalez2023,Lei2024,takagi-2024}. Remarkably, distinct perturbative orders occur exclusively in type III groups where opposite-spin sublattices are connected by fourfold rotations. Here, $\mathbf{M}_O$ remains at first order while $\mathbf{M}_S$ is suppressed to second order. This unconventional hierarchy serves as the microscopic cornerstone of type III altermagnets, identifying them as exhibiting a large AHE while maintaining a negligible net magnetic moment. Although $\rm RuO_2$ stands as the most prominent type III candidate, its magnetic structure remains controversial \cite{Berlijn2017, Hiraishi2024, keler-2024}, experimental verification of the AHE in other type III systems is still lacking.

The order disparity in type III groups originates from the joint constraints of the spin-only group $^{\infty m}1$ and the operation $\{-1\|(2l)\}\, (l = 1,2,3)$, where a $(2l)$-fold rotation connects opposite-spin sublattices. In the absence of spin-only group, the $\{-1\|(2l)\}$ symmetry dictates that $\mathbf{M}_O$ and $\mathbf{M}_S$ emerge at the $(l-1)$-th and $l$-th perturbation orders, respectively, rendering $\mathbf{M}_O$ always lower than $\mathbf{M}_S$. However, the presence of $^{\infty m}1$ modifies this relation depending on the parity of $l$. For type I and II groups with odd $l$, the collinear spin-only group impose additional symmetry constraints that push the onset of $\mathbf{M}_O$ from $(l-1)$-th to $l$-th order, thereby eliminating the order disparity. In contrast, for type III groups with even $l$, this disparity survives, maintaining $\mathbf{M}_O$ at the first order and $\mathbf{M}_S$ at the second order ($l=2$). Detailed derivations are provided in the Supplemental Material S2 \cite{suppl}.

\paragraph{Directional relation beyond MLG--}

\begin{figure}[htbp]
  \centering
  \subfloat[$\mathbf{M}_O,\mathbf{M}_S=0$\label{subfig:zero}]{%
    \includegraphics[width=0.2\textwidth]{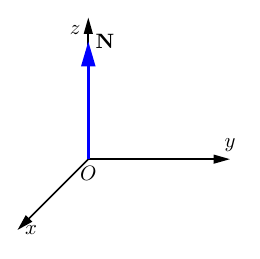}%
  }
  \subfloat[$\mathbf{M}_O,\mathbf{M}_S\parallel z$\label{subfig:paraz}]{%
    \includegraphics[width=0.2\textwidth]{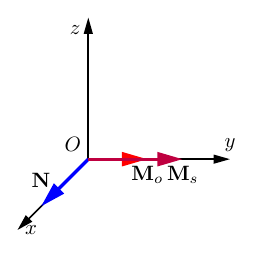}%
  }

  \subfloat[$\mathbf{M}_O,\mathbf{M}_S\parallel \mathbf{N}$\label{subfig:paran}]{%
    \includegraphics[width=0.2\textwidth]{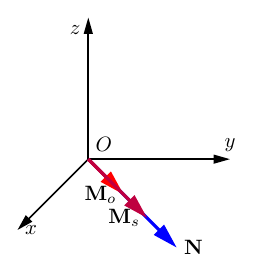}%
  }
  \subfloat[Arbitrary Directions\label{subfig:arb}]{%
    \includegraphics[width=0.2\textwidth]{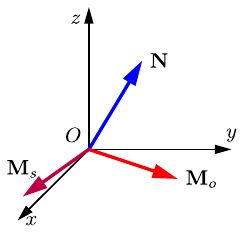}%
  }
  \caption{Directional evolution of spin and orbital magnetizations as a function of $\mathbf{N}$. (a) Disappearance of net magnetization when $\mathbf{N}$ aligns with a pure lattice-space rotation axis. (b) Typical alignment with the induced net magnetization perpendicular to $\mathbf{N}$. (c) Distinctive configuration in type III groups, where $\mathbf{M}_O$, $\mathbf{M}_S$ and $\mathbf{N}$ become mutually parallel. (d) Vanishing of the $z$-component of orbital magnetization even when $\mathbf{N}$ aligns along low-symmetry orientation.}
  \label{fig:directions}
\end{figure}

We further delineate the directional relation of $\mathbf{M}_O$ and $\mathbf{M}_S$ as a function of $\mathbf{N}$. First, whenever $\mathbf{N}$ aligns with the rotation axis of a pure lattice-space operation $\{1||R\}$, the spin Laue group is broken by SOC into a MLG that forbids the net magnetization [Fig. \ref{fig:directions}(a)]. This orientation-dependent rule establishes an explicit geometric criteria to identify the emergence of SOM within OSLGs. Beyond this, the induced net magnetization often remains perpendicular to $\mathbf{N}$. For example, in type I and type II groups, the collinear spin-only group constrains the $\mathbf{M}_S$ to be strictly perpendicular to $\mathbf{N}$, whereas $\mathbf{M}_O$ maintains perpendicular to $\mathbf{N}$ only when $\mathbf{N}$ aligns with specific high-symmetry directions [Fig. \ref{fig:directions}(b)]. 

In sharp contrast, in type III groups, particularly the $^{-1}4/^{1}m^{-1}m^{1}m^{\infty m}1$, when $\mathbf{N} \parallel [100]$, we find $\mathbf{M}_O^{(1)} \parallel \mathbf{M}_S^{(2)} \parallel [010]$, which is perpendicular to $\mathbf{N}$. As $\mathbf{N}$ rotates from $[100]$ to $[110]$, $\mathbf{M}_O$ and $\mathbf{N}$ maintain a mirror-symmetric relationship with respect to the $[110]$ axis, yielding a parallel alignment ($\mathbf{M}_O^{(1)} \parallel \mathbf{M}_S^{(2)} \parallel \mathbf{N} \parallel[110]$) [Fig. \ref{fig:directions}(c)]. Physically, this alignment stems from the SOC-induced symmetry breaking between opposite-spin sublattices. The presence of SOC eliminates all sublattice-interchanging operations within the corresponding MLG. Although a nonzero net magnetization is allowed, the residual symmetry strictly enforces its parallel alignment with $\mathbf{N}$. Consequently, this specific N\'{e}el configuration forces both orbital and spin magnetizations to align coaxially with $\mathbf{N}$, giving rise to the coaxial Hall effect. Unlike conventional cases with perpendicular $\mathbf{N}$ and net magnetization, this collinear relation eliminates multi-state degeneracy and switching barrier competition. A weak magnetic field enables a deterministic $180^\circ$ reversal of the N\'{e}el vector, similar to the switching in ferromagnets, thereby providing an efficient route for high-efficiency antiferromagnetic devices. 

Along low-symmetry orientations in type I (except monoclinic) and type III groups, the $\{-1||4^+_{001}\}$ or $\{1||R_{001}\}$ combined with $\{-1||m_{100}\}$ symmetry constrains the $\mathbf{M}_{O,z}^{(1)}$ to vanish, although the corresponding MLG permits a nonzero magnetization components along $z$ direction [Fig. \ref{fig:directions}(d)]. A more pronounced constraint arises in the $^{1}{-3}/^{-1}m ^{\infty m}1$ group. When $\mathbf{N}$ aligns with high-symmetry direction such as $[100]$, the corresponding MLG $m'$ allows net magnetization components along $y$ and $z$ directions. However, the symmetry still restricts $\mathbf{M}_{O,z}$ to higher-order SOC perturbations. Consequently, the anomalous Hall conductivity $\sigma_{xy}$ originates from higher-order SOC effects. Transport measurements should therefore focus on detecting the more prominent $\sigma_{xz}$ or $\sigma_{yz}$ components. A comprehensive visualization of the directional relation for all ten OSLGs is provided in the Supplemental Material S5 \cite{suppl}.

\begin{figure}[t]
  \centering
  \includegraphics[width=1\linewidth]{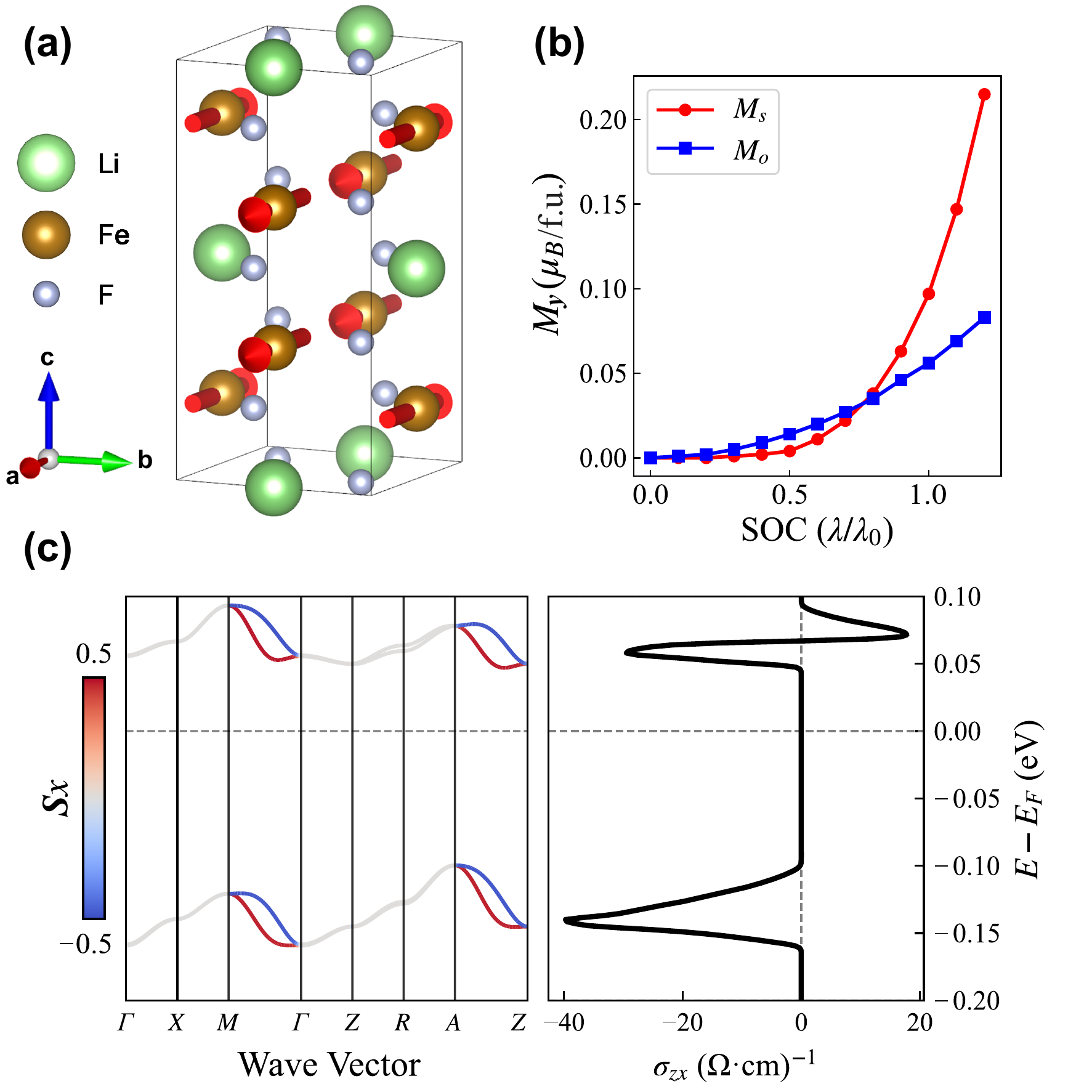}
  \caption{(a) Magnetic structure of LiFe$_2$F$_6$; (b) $\mathbf{M}_O$ and $\mathbf{M}_S$ as a dependence of SOC, where The horizontal axis represents the ratio of the SOC coefficient used in the calculation ($\lambda$) to the actual SOC value for LiFe$_2$F$_6$ ($\lambda_0$).; (c) $S_x$ projection band structure with SOC and anomalous Hall conductivity ($\sigma_{zx}$).}
  \label{fig:LFF}
\end{figure}

\paragraph{Decoupled scaling hierarchy in $\text{LiFe}_2\text{F}_6$ --}
We apply such a theoretical approach to evaluate the spin and orbital magnetizations of altermagnet $\text{LiFe}_2\text{F}_6$ [Fig.~\ref{fig:LFF}(a)] with an in-plane $\mathbf{N} \parallel [100]$. This magnetic structure possess the oriented spin space group $P^{-1}4_{2}/^1m^{-1}n^1m^{\infty_{100}m}1$. In the presence of SOC, this oriented spin space group is broken into the magnetic point group $m'mm'$, which allows a nonzero net magnetization along the [010] direction. Under symmetry constraints for tensors, both $\mathbf{M}_S$ and $\mathbf{M}_O$ vectors lie along the $[010]$ direction. Notably, this spin group enables the separation of the possible orbital and spin contributions to the magnetization, which are not distinguished by the magnetic group. The important insight for $\text{LiFe}_2\text{F}_6$ is that the SOC-induced magnetization of orbital origin, $\mathbf{M}_O$, is proportional to $\lambda$, whereas $\mathbf{M}_S$ exhibits a $\lambda^{2}$ dependence. This different polynomial dependence on SOC strength is confirmed by our density functional theory (DFT) calculations [Fig.~\ref{fig:LFF}(b)]. Specifically, despite consisting of only light elements, $\text{LiFe}_2\text{F}_6$ exhibits a sizable anomalous Hall conductivity up to 40 S/cm [Fig.~\ref{fig:LFF}(c)]. This underscores the advantages of type III altermagnets for spintronic applications.

\begin{figure}[t]
    \centering
    \includegraphics[width=1\linewidth]{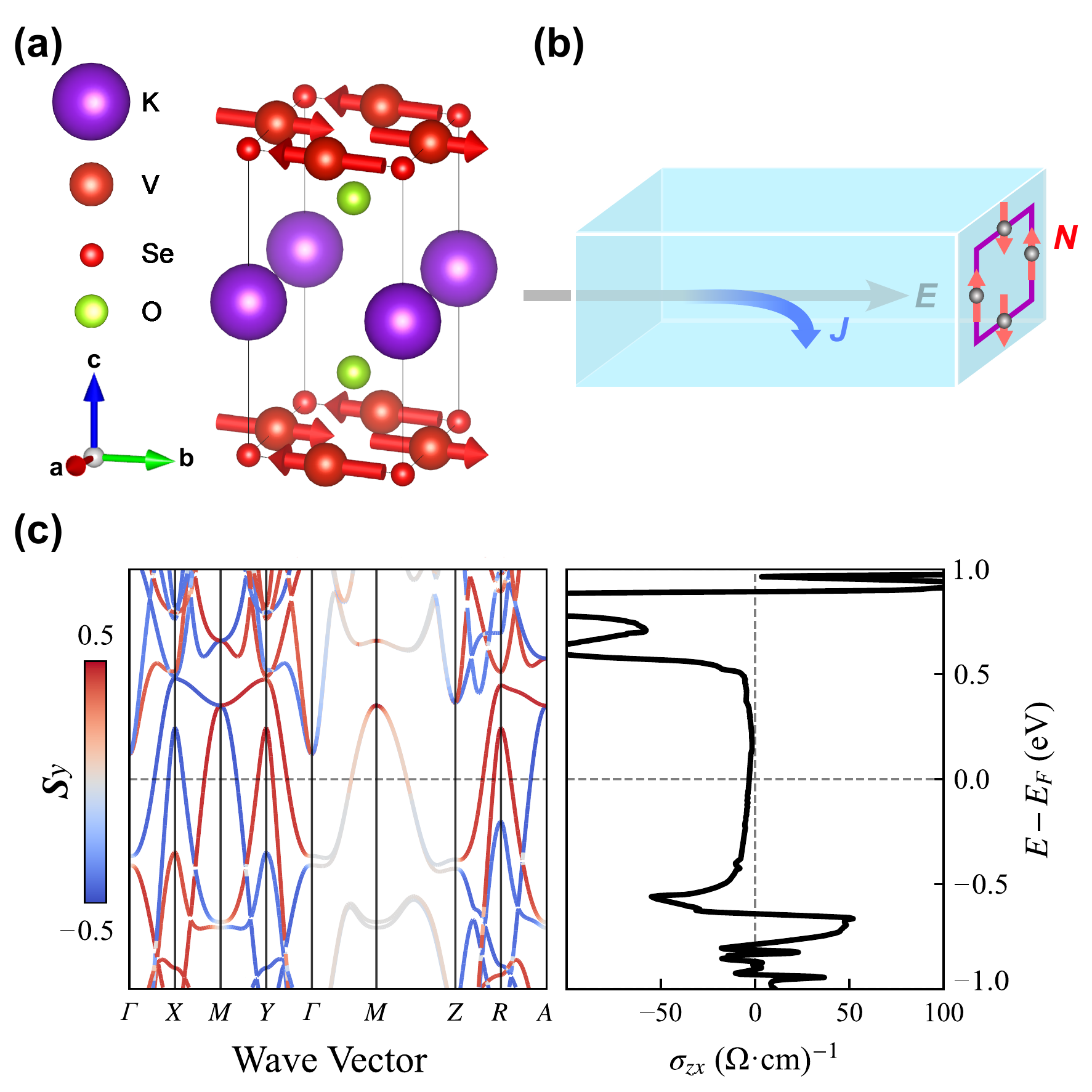}
    \caption{(a) Magnetic structure of KV$_2$Se$_2$O; (b) Schematic of coaxial Hall effect.; (c) $S_y$ projection band structure with SOC and anomalous Hall conductivity.}
    \label{fig:KVSO}
\end{figure}

\paragraph{Coaxial Hall effect--}
Based on the aforementioned symmetry principles, we perform a high-throughput screening of all altermagnets in the MAGNDATA database  \cite{Gallego20161,Chen2025}, uncovering 7 candidate SOMs that possess the key $\{-1||4\}$ symmetry. DFT calculations reveal that while five of them are large-gap insulators, two exhibit metallic behavior. Remarkably, the above-room-temperature altermagnet $\text{KV}_2\text{Se}_2\text{O}$  \cite{Jiang2025} emerges as a ideal platform to realize the coaxial Hall effect [Fig.~\ref{fig:KVSO}(a)]. Its magnetic structure can be described with oriented spin space group $P^{-1}4/^1m^{1}m^{-1}m^{\infty_{010} m}1$, which corresponds to $m'mm'$ under SOC. Note that due to a 45$^{\circ}$ shift in the crystallographic seeting of this oriented spin space group relative to the standard OSLG, the coaxial Hall effect can be realized when the $\mathbf{N}$ aligns along either the [100] or [010] axis. This effect fundamentally alters the operational principles of an antiferromagnetic anomalous Hall device [Fig.~\ref{fig:KVSO}(b)]. While most altermagnets exhibit an anomalous Hall vector perpendicular to $\mathbf{N}$, yielding a $V \propto \cos\psi$ angular dependence (where $\psi$ is the angle between $\mathbf{N}$ and the electric field $\mathbf{E}$), the coaxial Hall effect in $\text{KV}_2\text{Se}_2\text{O}$ enforces a $V \propto \sin\psi$ dependence. Consequently, the transverse Hall response vanishes identically when $\mathbf{E} \parallel \mathbf{N}$, but reaches its maximum when $\mathbf{E} \perp \mathbf{N}$, generating a Hall signal along the axis mutually orthogonal to both $\mathbf{E}$ and $\mathbf{N}$. In the absence of SOC, the system hosts a giant momentum-dependent spin splitting exceeding $1$ eV along the $\Gamma$-$X$ path, exhibiting a prototypical $d$-wave spin texture that confirms its altermagnetic nature (see Supplementary Material S7 \cite{suppl}). Crucially, although the introduction of SOC leaves the overall band structure nearly unchanged, it generates substantial Berry curvature at band crossings, yielding a sizable anomalous Hall conductivity of $\sim 100$ S/cm near the Fermi level despite a negligible net magnetization of less than $10^{-3}\,\mu_B$.

\paragraph{Summary--}

Our approach uncovers the SOC dependence of spin and orbital magnetizations in SOMs, thereby clarifying the puzzling relationship between net magnetization and the anomalous Hall effect in altermagnets. Traditional magnetic group analysis inherently coupled these two magnetizations, obscuring a core physical picture: the AHE related by $\mathbf{M}_O$ can be completely decoupled from the stray-field-related $\mathbf{M}_S$. This decoupling reveals that the SOC-driven magnetism in SOM is not a monolithic entity. In type III altermagnets, this decoupling manifests as a symmetry-protected disparity in perturbation orders, maximizing the AHE signal while keeping net magnetization minimal. Meanwhile, the emergent coaxial Hall effect yields a unique transverse voltage profile and enables deterministic 180° reversal of $\mathbf{N}$ under weak magnetic fields, thereby achieving ferromagnetic-like operational simplicity within an altermagnet. Furthermore, this framework can be generalized to non-collinear magnets, where the rule that $\mathbf{M}_O$ and $\mathbf{M}_S$ differ by at most one perturbative order in SOC still holds. This tensor approach holds broad promise for describing magnetic anisotropy, electric polarization, and quantum transport, such as the separation of spin and orbital current, establishing a comprehensive framework for the symmetry-guided design of next-generation quantum materials.

\begin{acknowledgments}
This work was supported by National Key R$\&$D Program of China under Grant No. 2025YFA1411300, National Natural Science Foundation of China under Grants No. 12525410, No. 12274194, No. 12574275, and No. 12534003, and Guangdong Provincial Quantum Science Strategic Initiative under Grant No. GDZX2401002, Guangdong Provincial Key Laboratory for Computational Science and Material Design under Grant No. 2019B030301001, Shenzhen Science and Technology Program (Grant No. RCJC20221008092722009, No. 20231117091158001 and No. QNXMC20250701094702004), the Innovative Team of General Higher Educational Institutes in Guangdong Province (Grant No. 2020KCXTD001) and Center for Computational Science and Engineering of Southern University of Science and Technology.
\end{acknowledgments}

\bibliography{apssamp}

\end{document}